\documentclass[runningheads]{llncs}
\usepackage[T1]{fontenc}
\usepackage{comment}
\usepackage{url}
\usepackage{biblatex} 
\usepackage{listings}
\usepackage{float}
\usepackage{xcolor}
\usepackage{graphicx}
\usepackage{longtable}
\usepackage{amsmath}
\usepackage{array}
\usepackage{booktabs}
\usepackage{threeparttable}
\usepackage{makecell}
\usepackage{xcolor,mdframed}
\usepackage[export]{adjustbox}
\usepackage{graphicx}
\usepackage{subcaption}
\usepackage{wasysym}
\def\NAMEOFDATASET{KVP10k }
\def\NDOCS{10707}
\def\IOUTHRESH{0.3}
\def\EDITHRESH{0.2}

\lstdefinestyle{json-style}{
    stringstyle=\color{red},
    keywordstyle=\color{blue}
}
\lstdefinelanguage{json}{
    basicstyle=\normalfont\ttfamily,
    string=[s]{"}{"},
    stringstyle=\color{red},
    comment=[l]{:},
    commentstyle=\color{blue},
    keywordstyle=\color{blue}
}

\begin{document}

\title{\NAMEOFDATASET: A Comprehensive Dataset for Key-Value Pair Extraction in Business Documents}
\author{Oshri Naparstek\inst{1} \and
Ophir Azulai\inst{1} \and
Inbar Shapira\inst{1} \and
Elad Amrani\inst{1} \and
Yevgeny Yaroker\inst{1} \and
Yevgeny Burshtein\inst{1} \and
Roi Pony\inst{1} \and
Nadav Rubinstein\inst{1} \and
Foad Abo Dahood\inst{1} \and
Orit Prince\inst{1} \and
Idan Friedman\inst{1} \and
Christoph Auer\inst{2} \and
Nikolaos Livathinos\inst{2} \and
Maksym Lysak\inst{2} \and
Ahmed Nassar\inst{2} \and
Peter Staar\inst{2} \and
Udi	Barzelay\inst{1}}
\authorrunning{O. Naparstek et al.}
%
\institute{IBM Research Israel, University of Haifa Campus, Mount Carmel, Haifa 3498825, Israel\\
IBM Research Zurich,Säumerstrasse 4, 8803 Rüschlikon, Switzerland
}



\maketitle
\begin{abstract}
In recent years, the challenge of extracting information from business documents has emerged as a critical task, finding applications across numerous domains. This effort has attracted substantial interest from both industry and academy, highlighting its significance in the current technological landscape. Most datasets in this area are primarily focused on Key Information Extraction (KIE), where the extraction process revolves around extracting information using a specific, predefined set of keys. Unlike most existing datasets and benchmarks, our focus is on discovering key-value pairs (KVPs) without relying on predefined keys, navigating through an array of diverse templates and complex layouts. This task presents unique challenges, primarily due to the absence of comprehensive datasets and benchmarks tailored for non-predetermined KVP extraction. To address this gap, we introduce \NAMEOFDATASET, a new dataset and benchmark specifically designed for KVP extraction. The dataset contains \NDOCS $ $ richly annotated images. In our benchmark, we also introduce a new challenging task that combines elements of KIE as well as KVP in a single task. \NAMEOFDATASET sets itself apart with its extensive diversity in data and richly detailed annotations, paving the way for advancements in the field of information extraction from complex business documents. 

\end{abstract}

\section{Introduction}
Extracting KVPs from business documents is a critical task that holds significant importance for businesses today. In an increasingly data-driven world, organizations generate and receive vast amounts of unstructured textual data in the form of invoices, contracts, reports, and other documents. The ability to efficiently extract relevant information in the form of key-value pairs from these documents can greatly benefit businesses. It not only streamlines data entry processes but also enables quick and accurate access to essential information, leading to improved decision-making, enhanced efficiency, and better overall business operations. In this context, KVP extraction plays an important role in transforming unstructured data into actionable insights, helping companies stay competitive in their respective industries.

In exploring the landscape of information extraction from documents, it is essential to distinguish among KIE, Document Question Answering (DQA), and KVP extraction. These tasks, while related, diverge significantly in their objectives and methodologies.

KIE, as the most established of the trio, focuses on categorizing text snippets into a predefined set of classes. This process often involves the aggregation of related textual entities under unified labels, making it a task of entity recognition and classification at its core. The simplicity of KIE, relative to the other tasks, stems from its reliance on a known set of classes, reducing the complexity to the identification and classification of text according to these categories. Representative datasets in this domain include  CORD\cite{park2019cord}, SROIE\cite{huang2019icdar2019SORIE}, Kleister-NDA\cite{stanislawek2021kleister}, VRDU\cite{wang2023vrdu}, Kleister-Charity\cite{stanislawek2021kleister}, and EPHOIE\cite{wang2021towards}.

Another related task is Document Question Answering. This task introduces a different paradigm, where the task is not to classify text into predefined categories but to locate and extract answers to user-posed questions directly from the text. This task eliminates the need for a fixed set of labels, instead requiring the model to understand the question's intent and retrieve relevant information from the document. The dynamic nature of the questions introduces variability and complexity, as the model must adapt to the diverse range of inquiries. DocVQA\cite{mathew2021docvqa} is a notable dataset in this field, challenging models with a wide array of question-answer scenarios.

KVP extraction presents challenges akin to those encountered in document-based question answering, largely due to the absence of predefined keys. In question answering, complexity often arises when synthesizing an answer necessitates aggregating information from various sections within a document. Despite this, the objective remains to distill a singular, precise answer. In contrast, KVP extraction demands a comprehensive retrieval of all pertinent keys and values scattered throughout a document, expanding the scope beyond seeking a specific answer to encompass a broader extraction task. This task demands an understanding of document structure and content to discern the relationships between different pieces of information, often dealing with hierarchical key-value structures. Datasets such as FUNSD\cite{jaume2019funsd} and XFUND\cite{xu2022xfund} are examples of the complexities of KVP extraction, presenting diverse documents where models must infer and extract a broad spectrum of information without relying on a fixed schema.

Despite the comprehensive and demanding nature of KVP extraction, a notable challenge within this domain is the current state of available datasets. The existing datasets for KVP extraction, such as FUNSD\cite{jaume2019funsd} and XFUND\cite{xu2022xfund}, are somewhat limited in scope and diversity. They tend to be smaller in size, which can restrict the depth of training and the robustness of models developed using these resources. Even newer datasets such as Form-NLU\cite{ding2023form} and SIBR\cite{yang2023modelingSIBR} are only 857 and 1,000 pages respectively. Furthermore, these datasets often lack variety in their sources, presenting a narrow view of the potential applications and scenarios where KVP extraction could be applied. This limitation in dataset quality and diversity poses significant challenges for researchers and practitioners aiming to develop models that are capable of performing well across a wide range of real-world documents and contexts. The need for larger, more varied datasets is crucial in pushing the boundaries of what KVP extraction models can achieve, ensuring they are versatile and effective across diverse document types and industries.



In recent years, there has been a growing interest in the domain of key information extraction and key-value pair extraction from various types of documents and many models were developed for the task of document understanding\cite{huang2022layoutlmv3}\cite{lee2023formnetv2}\cite{hong2020bros}\cite{mathur2023layerdoc}\cite{wang2020docstruct}\cite{hwang2020spatial}\cite{perot2023lmdx}. This growing enthusiasm is reflected in both academic research and industry applications, where the need to automate the extraction of critical data from documents such as legal contracts and medical records is increasingly apparent. However, despite this surge in interest and the evident practical importance of this task, a noticeable gap remains in the field: the absence of a comprehensive and high-quality dataset tailored specifically for key-value pair extraction from documents. This notable void has underscored the necessity for collaborative efforts within the research community to address this deficiency and create a resource that can significantly advance state-of-the-art document analysis and information extraction, ultimately benefiting a wide array of businesses and organizations across diverse domains.
 
In response to the gap in Key-Value Pair (KVP) extraction from documents, we introduce \NAMEOFDATASET. This dataset is distinguished by its comprehensive scope and focus on KVP extraction. Our contributions through this work are threefold:

\begin{enumerate}
    \item {\bf New Dataset} -- \NAMEOFDATASET includes \NDOCS $ $  pages, making it the largest dataset available for KVP extraction. It features a broad array of keys and precise annotations, with text labeled as keys or values, providing a solid basis for training and evaluation.
    \item {\bf New Benchmark with Metrics} -- We present a benchmark for KVP extraction, offering a framework for model comparison and performance evaluation. This facilitates a clearer understanding of model capabilities in the field.
    \item {\bf Baseline Results} -- Initial baseline results are shared to establish a foundation for subsequent research, aiming to enhance KVP extraction methods.
\end{enumerate}
\NAMEOFDATASET aims to support the advancement of document processing technologies by providing high-quality data and a platform for rigorous research.

\section{Related work}
In this work, we have considered a wide range of existing datasets that contribute to the fields of KVP extraction and KIE. Notably, FUNSD\cite{jaume2019funsd} and XFUND\cite{xu2022xfund} are prominent datasets for KVP tasks. Additionally, the SIBR\cite{yang2023modelingSIBR} dataset is relevant for KVP tasks, particularly focusing on camera-captured image scenarios with real-world complexities such as blur, noise, and uneven illumination (in the wild). However, it is important to recognize that these datasets are relatively small and typically classify entities of KVPs as either questions or answers, without annotating common class types like dates or names. This limitation restricts their applicability for more complex or nuanced applications. In our endeavor, we have enhanced this foundation by annotating both KVPs and KIE elements, introducing a more detailed classification system with 17 distinct classes.

Turning our attention to KIE, datasets such as CORD\cite{park2019cord}, SROIE\cite{huang2019icdar2019SORIE}, Kleister-NDA\cite{stanislawek2021kleister}, VRDU\cite{wang2023vrdu}, Kleister-Charity\cite{stanislawek2021kleister}, and EPHOIE\cite{wang2021towards} have provided valuable insights and benchmarks. Yet, a common limitation among these resources is their size, which often restricts the depth and breadth of research and application development. Moreover, many of these datasets are derived from specific, homogeneous sources or templates, which may not fully represent the diversity and variability encountered in real-world data.

\begin{figure}[htbp]
\centering
\includegraphics[width=\columnwidth]{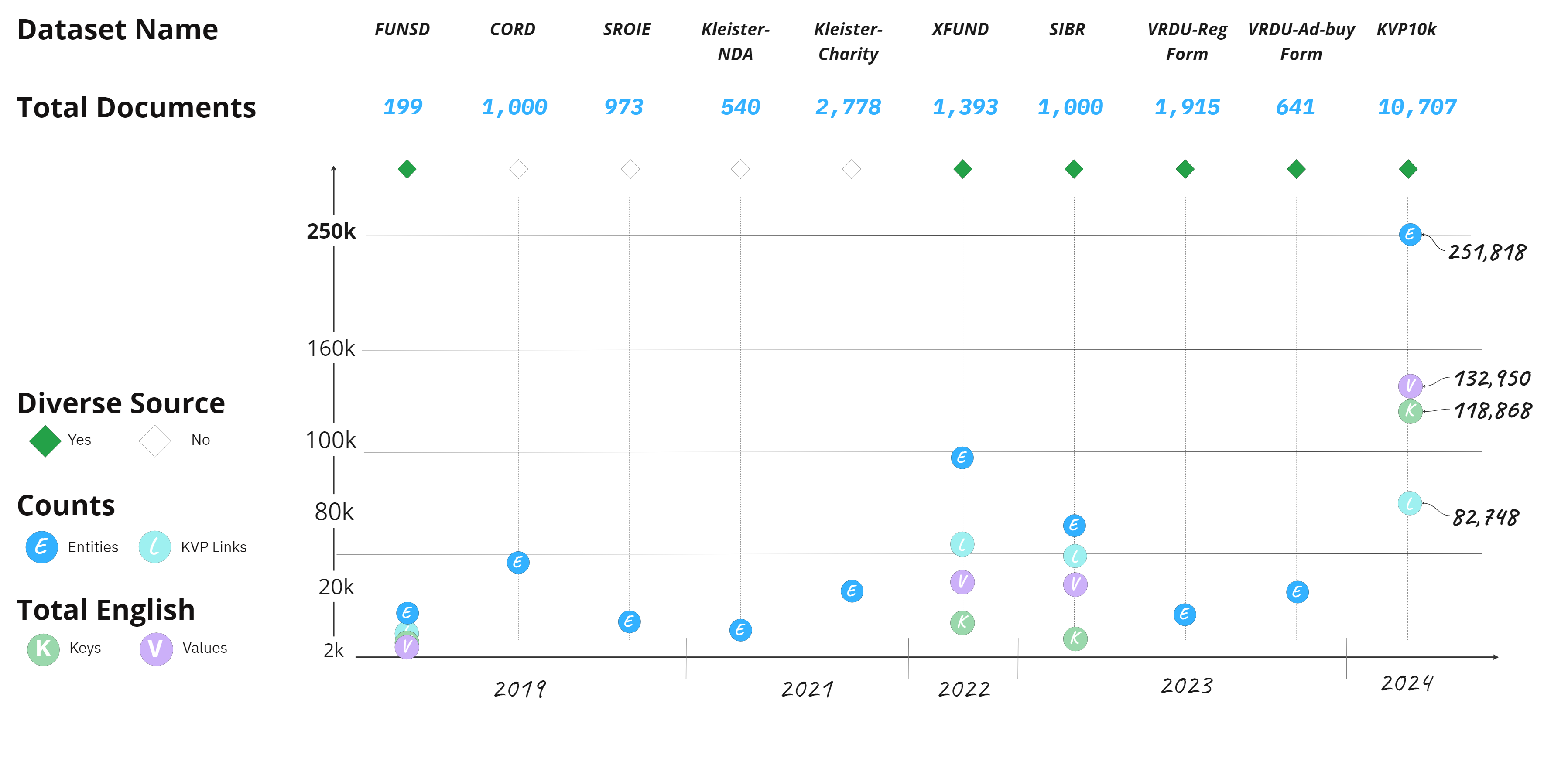}
\caption{Comparative overview of \NAMEOFDATASET versus other datasets: Comparing the Number of Documents, Entities, Keys, Values, and Links}
\label{fig:dataset_graph}
\end{figure}

In contrast, our dataset has been meticulously compiled to ensure a broad representation of sources and templates. This diversity is crucial for developing robust models capable of handling the wide range of formats and contexts in which KVP and KIE tasks are applied. Furthermore, unlike XFUND, which relies on synthetic data, our dataset is composed entirely of real-world documents. This choice underlines our commitment to authenticity and applicability, ensuring that the insights and models derived from our dataset are useful in real-world scenarios. Detailed comparison is provided in Fig \ref{fig:dataset_graph}.

In summary, while existing datasets have laid important groundwork in the fields of KVP and KIE, our dataset seeks to address some of their key limitations by offering a larger, more detailed and diverse collection of real-world data. We hope that this contribution will not only support the advancement of current research but also inspire new directions in the extraction and understanding of key information from complex documents.

\section{Data Acquisition}

Our data acquisition process leveraged two primary sources to ensure a diverse and comprehensive dataset: extensive web data from Common Crawl and a collection of images from publicfiles.fcc.gov.

From Common Crawl, we employed a systematic approach to download indices and identify URLs of PDFs, focusing on over 40,000 targeted domains. These domains encompassed a broad spectrum of sources, including various companies, government bodies, and educational institutions. Applying filters to these URLs based on a pre-defined list of relevant and reliable domains, we efficiently collected a vast dataset pertinent to our research needs.

\begin{figure}[h]
\centering
\includegraphics[width=1\columnwidth]{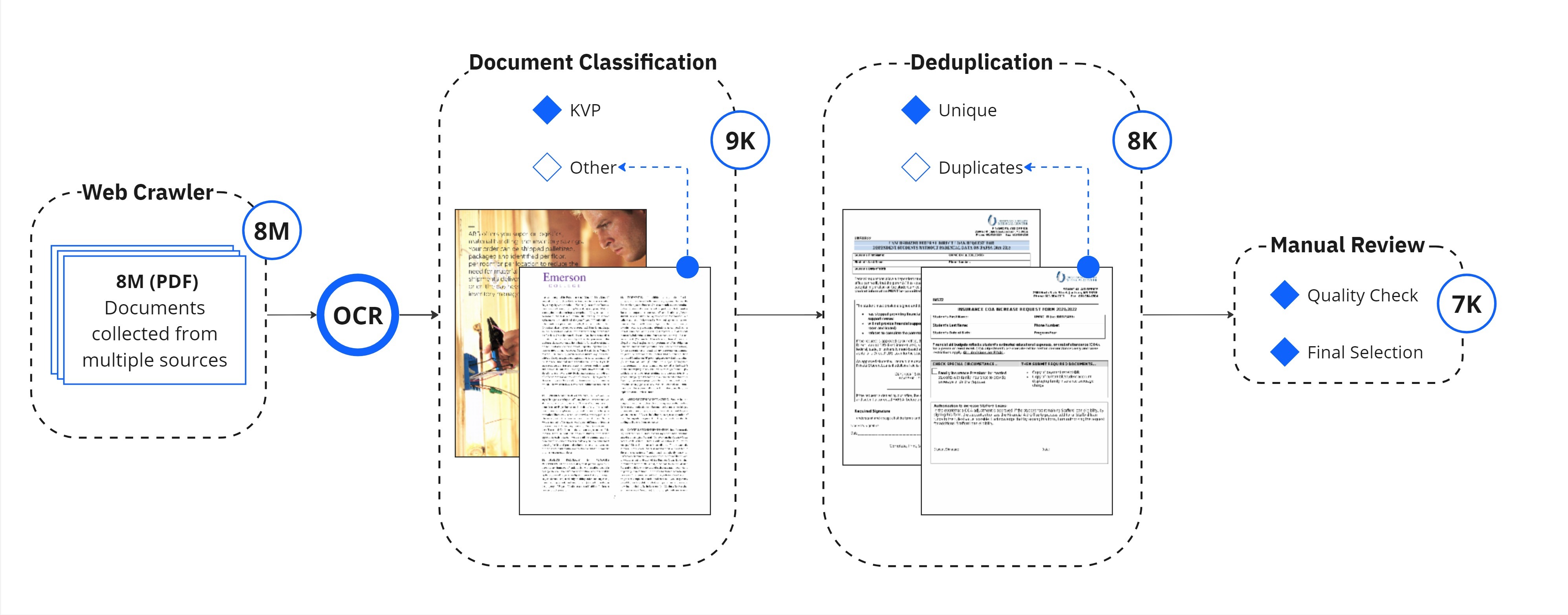}
\caption{A schematic describing the data collection process using web crawling.}
\label{fig:data_acq}
\end{figure}

In the data filtering phase, we chose a subset of 8 million documents from the initially extracted web data. For categorizing these documents, we created a classifier to distinguish between documents suitable for KVP extraction and the rest. This classifier was developed employing a Longformer model\cite{beltagy2020longformer}, combined with a Roberta tokenizer(2019)\cite{liu2019roberta}. Utilizing the 'allenai/longformer-base-4096' architecture as a foundation. The classifier was trained for binary classification over 20 epochs with a learning rate of 3e-4, employing the Adam optimizer. This training process used a dataset comprising 2,378 documents suitable for KVP extraction and 12,610 documents classified as non-KVP from the 8-million-document subset. The classifier's performance was notable, achieving a precision of 0.97, a recall of 0.55, and an F1-score of 0.7, culminating in an overall accuracy of 0.92. We also engineered two distinct rule-based classifiers. The primary classifier targets documents containing more than a specified number of words, N, which include pre-established substrings frequently observed in business documents such as "bill," "ship," "total," "sub," etc. Our secondary classifier is tailored to identify documents featuring more than K independent clauses, concluding with a colon. It then scans for an ensuing independent clause either directly after the colon or in the subsequent line.
A schematic describing the data acquisition process for the common crawl data is shown in Fig.\ref{fig:data_acq}.

Besides the 7,000 images, we also acquired additional pages from publicfiles.fcc.gov by retrieving the initial 44,000 PDF links. Out of these, we randomly chose 5,000 PDFs, from which we then randomly selected between 1 to 5 pages per PDF. 

To maintain diversity in the data and mitigate potential biases, a deduplication filter was employed, leveraging OCR\cite{naparstek2022businet} and sentence outputs. A criterion was set where documents were deemed alike if they had a minimum overlap of six sentences, with each sentence comprising at least three words. This deduplication method was applied to the document batch processed in the second phase, organizing them into clusters. Within each cluster, a single document was chosen at random. 

In total, we gathered 3,524 pages from publicfiles.fcc.gov and 7,183 pages from Common Crawl.

\section{Data Annotation}
In the process of data annotation, a set of specific guidelines was established to ensure consistency and accuracy. These guidelines cover various annotation types, each with its own defined characteristics and purpose. The guidelines were designed to cater to different elements within the documents. 
The annotation process is structured to categorize and link textual elements within a given task. This involves enclosing relevant text within a defined area and assigning an appropriate label from a pre-determined set of label types. These labels include 'Text', and 'Handwriting-text'. Subsequently, a linkage is established from a value to its corresponding key. 

In addition we introduce two label types without linking,  'unvalued-key' and unkeyed-values.
Moreover, for unkeyed-values, a range of label types is available to categorize them appropriately. These types encompass 'Floating Document Type', 'Floating Document Title', 'Floating Year', 'Floating Date', 'Floating Name', 'Floating Address', 'Floating Phone', 'Floating Email', 'Floating Website', 'Floating Amount', and 'Floating Text' for instances that do not align with the aforementioned categories. This systematic approach ensures clarity and coherence in the annotation of the document, facilitating a structured and comprehensible representation of the data. An example for an annotated page is given in Fig \ref{fig:telus1}. Examples of the annotation format are given in appendix \ref{appendix_annotation}.
\begin{figure}[htbp]
\centering
\begin{mdframed}[backgroundcolor=gray!50,linecolor=gray!50]
\includegraphics[width=1\columnwidth]{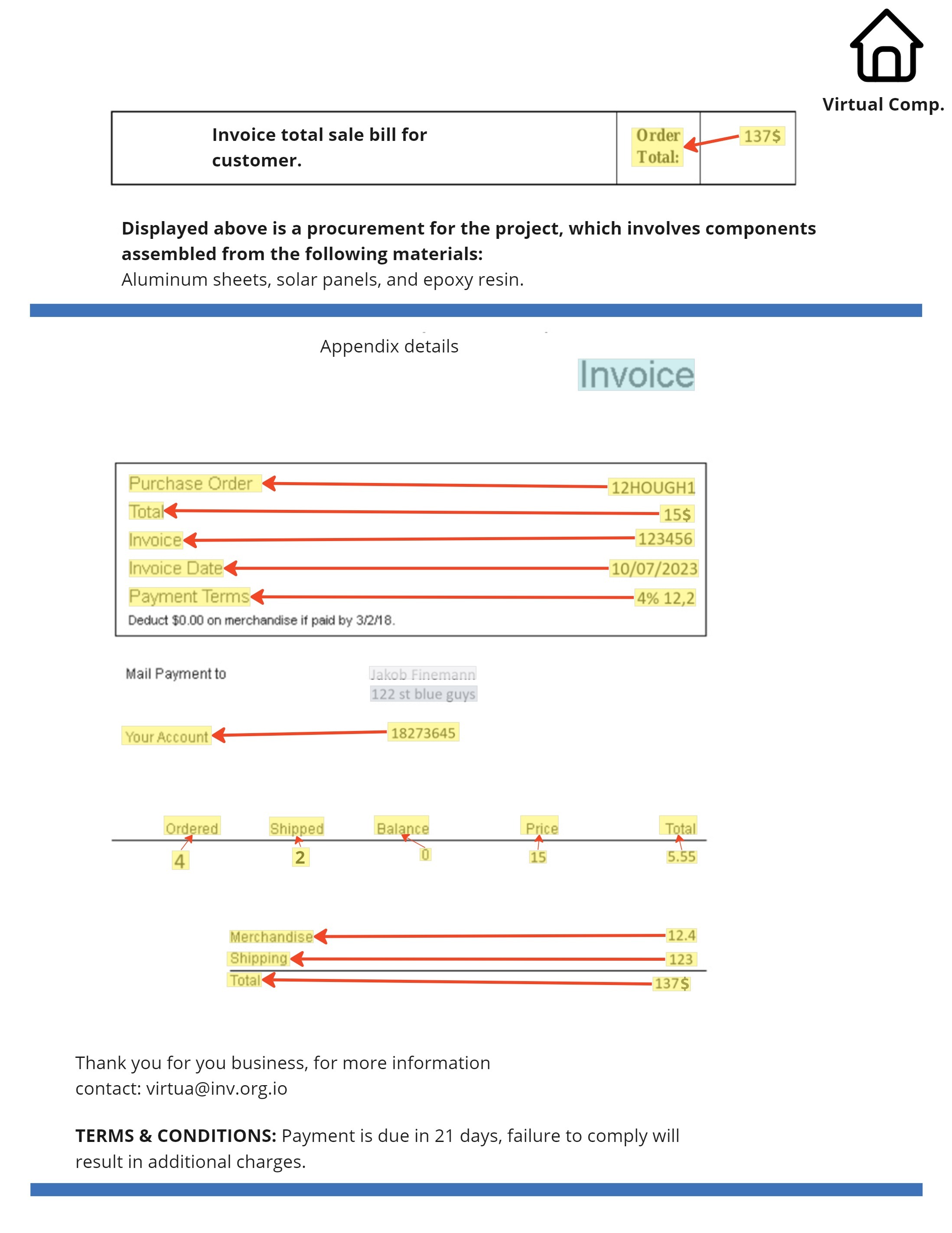}
\caption{Example of an annotated page}
\label{fig:telus1}
\end{mdframed}
\end{figure}

\section{Dataset Characteristics}
\subsection{Diverse Source of Images}
\begin{figure}[htbp]
\centering
\includegraphics[width=1\columnwidth]{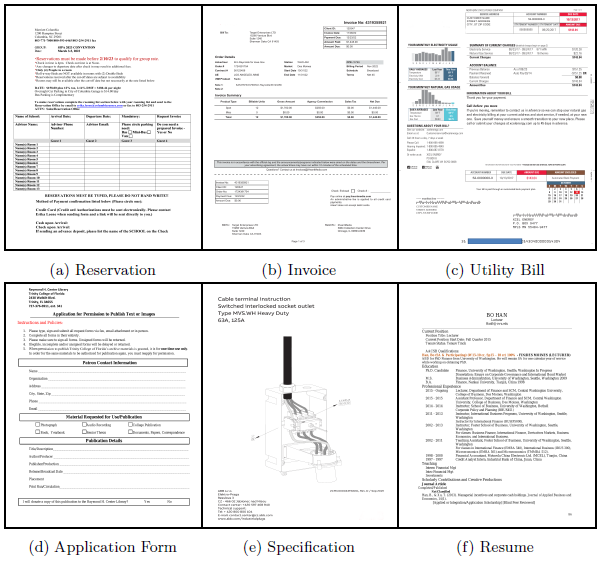}
\caption{Exemplifying Versatility: A collage of diverse document categories from \NAMEOFDATASET Dataset}
\label{fig:Exemplifying_Versatility}
\end{figure}

Our dataset goes beyond mere visual richness, providing a deep understanding of the text and how things relate to each other, setting new standards for how we understand documents, particularly in key-value pair extraction. This richness in text is not just about having different kinds of documents or complex designs. It's about exploring the detailed layers of text and the intricate ways text parts interact across various documents. Covering a wide range of document types, from business papers to scientific studies, \NAMEOFDATASET includes a variety of words and specialized terms. This range of documents helps in studying complex text patterns, meanings, and hints that help train models to understand not just what documents look like but also the following two properties: 1) what the information means, 2) why it is important. 
Fig. \ref{fig:Exemplifying_Versatility} presents a selection of images from different types of documents. These examples showcase the visual and text variety in \NAMEOFDATASET. In addition, they showcase the complexity of document designs and the thorough assignment of labels that help train deep learning models.

\subsection{Dataset Statistics}
\begin{figure}[htbp]
\centering
\includegraphics[width=1\columnwidth]{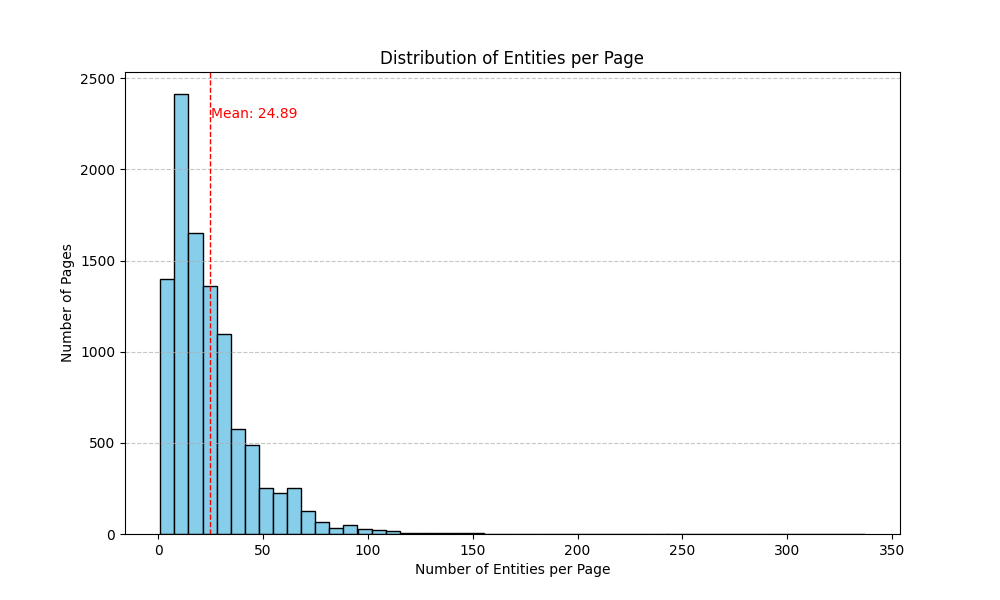}
\caption{Distribution of entities per page in \NAMEOFDATASET.}
\label{fig:entities_per_page_histogram}
\end{figure}

In this section, we present some dataset statistics to provide an understanding of the characteristics of \NAMEOFDATASET. The dataset is diverse and richly annotated, covering various document types and layouts.
In Figure \ref{fig:entities_per_page_histogram}, we present the distribution of entities per page in the dataset. 

\begin{figure}[htbp]
\centering
\includegraphics[width=1\columnwidth]{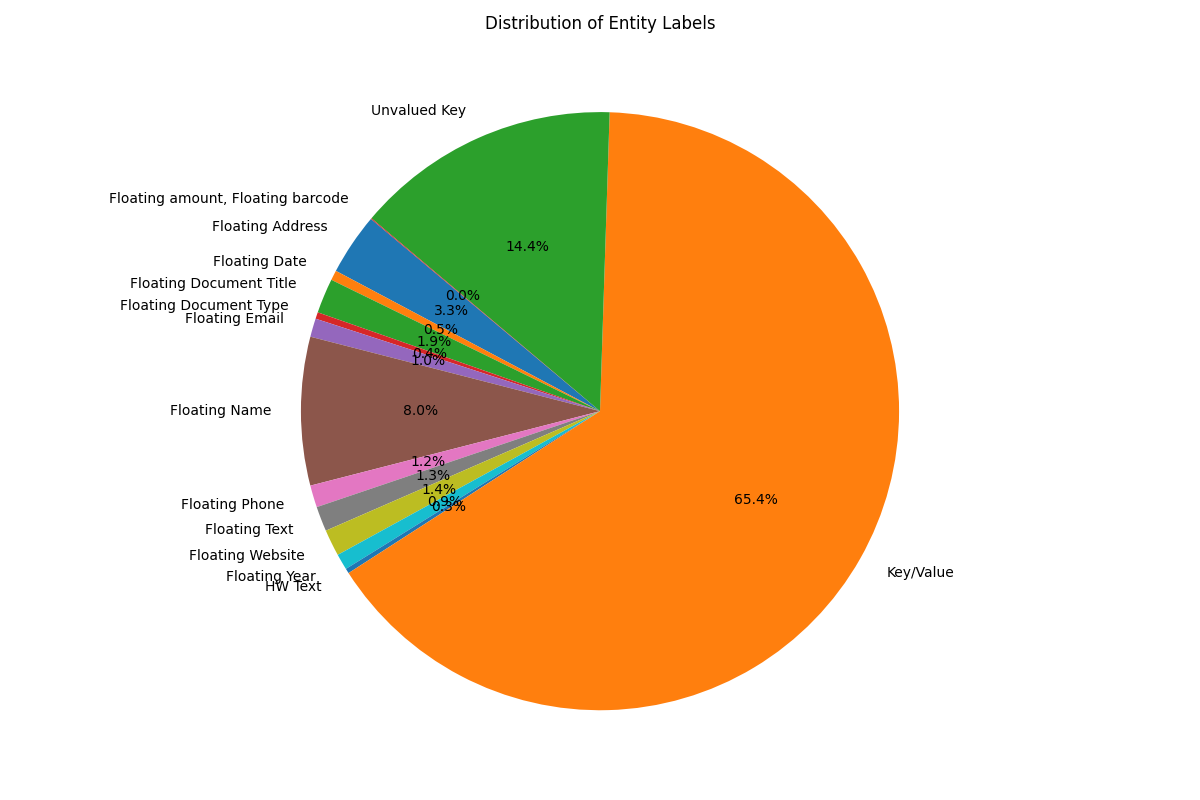}
\caption{Distribution of entity labels in \NAMEOFDATASET.}
\label{fig:entity_labels_pie_chart}
\end{figure}

Figure \ref{fig:entity_labels_pie_chart} provides an overview of the distribution of entity labels in the benchmark dataset. This pie chart shows the relative proportions of different entity labels present in the dataset, including labels such as floating name, text, phone, date, key/value, etc.

Again, Figure \ref{fig:dataset_graph} presents a comparative analysis of \NAMEOFDATASET against existing ones, highlighting our dataset's superior quantity of documents, entities, links, keys, and values.

Together, these statistics illustrate the diversity and complexity of this dataset, providing insights into its composition and structure. 

\section{Benchmark}
To support the community in evaluating KVP extraction systems, we have developed a comprehensive benchmarking tool. This utility is crafted to assess the performance of various KVP extraction models, providing a uniform and reliable method for researchers and developers to gauge the effectiveness of their solutions.

Our benchmarking code includes implementations of the metrics focusing on the location of an entity, the textual meaning of an entity, and a combined approach. It facilitates a nuanced assessment, shedding light on different facets of a model's performance. We have designed the code with user-friendliness in mind, ensuring it can be easily integrated with diverse KVP extraction models. This feature is particularly beneficial for researchers and practitioners in the domain, streamlining the process of evaluating and enhancing their KVP extraction tools.

The benchmark code is openly accessible and can be found at our GitHub repository\footnote{\url{https://github.com/IBM/KVP10k}}. We invite the community to use this resource to propel forward the development of more precise and efficient KVP extraction technologies.

Our goal with this benchmarking code is to establish a standardized approach for evaluating KVP extraction systems, thus promoting continuous progress and innovation in this area.

\subsection{Tasks}

This benchmark is designed to rigorously evaluate the performance of algorithms in the extraction of key-value pairs from documents. It consists of two distinct tasks, each tailored to assess specific aspects of the algorithms' capabilities.
\subsubsection{Entity Recognition Task}
Similar to FUNSD\cite{jaume2019funsd}, the primary objective of this task is to identify key and value entities within a document. The effectiveness of an algorithm is quantified through two metrics: normalized edit distance and Intersection Over Union (IOU). An entity is considered a successful 'hit' if it satisfies the following criteria:

\begin{itemize}
    \item The normalized edit distance from the ground truth is below \EDITHRESH.
    \item The IOU with the ground truth exceeds \IOUTHRESH.
\end{itemize}

The overall performance of the algorithm is assessed using the F1 score.


\subsubsection{Key-Value Pair Detection Task}
The second task extends the challenge by focusing on the detection of key-value pairs in the document, as well as identifying unkeyed values and unvalued keys. An unkeyed value refers to a value present without an explicit key, such as a date without a preceding "Date:" label. An unvalued key refers to a key present with empty value, such as in empty fillable form. This task also encompasses the detection of values with no associated keys. 

We have three  metrics for this task. Location only, text only and a combination of the two.
\begin{itemize}
    \item Location only metric: a key-value pair is considered a 'hit' if both the key and value have IOU above \IOUTHRESH.
    \item Text only metric: a key-value pair is considered a 'hit' if both the key and value has normalized edit distance below \EDITHRESH.
    \item Combined metric: a key-value pair is considered a 'hit' if both the key and the value have normalized edit distances below \EDITHRESH and have IOU above \IOUTHRESH.
\end{itemize}
Unkeyed values are handled as key-value pairs where the value is subject to the previously defined criteria, and the key is identified through an exact text match. Unlike the standard key-value pairs, the detection of the key in these instances does not require a threshold for Intersection over Union (IOU) to be considered a successful hit.
Unvalued keys are processed as key-value pairs where the key meets the established criteria and the value is explicitly recognized as empty. For these instances, the assessment of the value does not necessitate the Intersection over Union (IOU) threshold.

The evaluation metric remains consistent with the first task, utilizing the F1 score. The precision and recall are calculated based on two steps: 1) number of  correctly identified pairs, unkeyed values, and unvalued keys in relation to the total number of such entities in each image, and 2) then averaging the scores per image over the total number of images.

This task presents a novel integration of elements from both KIE and KVP tasks, offering a unique approach to document analysis. For unkeyed values, the task aligns with KIE tasks, akin to those seen in datasets like CORD\cite{park2019cord}. Conversely, when addressing key-value pairs, the methodology parallels entity detection and linking tasks, similar to what is observed in datasets such as FUNSD\cite{jaume2019funsd}. Combining the approaches allows for appropriate handling of unpredictable and diverse document information.

\subsection{Output Format}
The format we have adopted for presenting the results of our benchmark analysis is intentionally straightforward, designed for ease of interpretation and use. Specifically, the data output is organized as a sequence of dictionaries, with each dictionary entry comprising a pair of key-value elements. This structure captures not only the textual content but also the spatial positioning of each element, facilitating a comprehensive understanding of the data's layout.

In Listing~\ref{lst:output_format_unkeyed}, we provide a sample of the output format. This example illustrates the structure of the data, including the categorization of key-value pairs (`unkeyed`, `unvalued` or as a specific `kvp`) and the associated spatial information (`bbox`) for each text element, represented as coordinates on the document.

\begin{lstlisting}[float,language=json,style=json-style, breaklines=true, frame=single, caption={Illustrative example of the output format}, label=lst:output_format_unkeyed]
{
    "kvps_list": [
        {
            "type": "unkeyed",
            "key": {
                "text": "Example Key Text 0"
            },
            "value": {
                "text": "Associated Value Text 0",
                "bbox": [x1, y1, x2, y2]
            }
        },
        {
            "type": "unvalued",
            "key": {
                "text": "Example key text 1",
                "bbox": [x1, y1, x2, y2]
            },
            
        },
        {
            "type": "kvp",
            "key": {
                "text": "Example Key Text 2",
                "bbox": [x1, y1, x2, y2]
            },
            "value": {
                "text": "Associated Value Text 2",
                "bbox": [x1, y1, x2, y2]
            }
        }
    ]
}
\end{lstlisting}

In this format, each `key` and `value` is delineated by their respective textual content and a bounding box (`bbox`), the latter specifying the coordinates of the text's location on the page, thus offering a dual perspective on the data: textual and spatial. The `type` field distinguishes between `unkeyed` entries, where the value's text is provided and the key's text is set as the appropriate label from the unkeyed-values range of label types, `unvalued` entries, where only the key's text is provided, and `kvp` entries, where both key and value include spatial data. This nuanced approach ensures a richer dataset, conducive to more insightful analysis.




\subsection{Baselines}

In this baseline section, we delve into the preliminary outcomes of our exploration into the tasks outlined earlier, employing a strategy influenced by the LMDX framework \cite{perot2023lmdx}. Our initial step involves processing the OCR-derived text from each document, converting it into a format amenable to integration with a large language model. This process entails arranging the OCR\cite{smith2007overview} text in a systematic top-to-bottom and left-to-right order, where each text line is tagged with its bounding box coordinates, denoted as [x1,y1,x2,y2].

While considering various models for this task, we recognized certain limitations in encoder-based models like LayoutLMv3\cite{huang2022layoutlmv3}. These models primarily function as token classifiers, which inherently restricts their capability in performing tasks that require key-value pair linking, as they lack the mechanism to generate new tokens necessary for establishing these links. Moreover, their performance is significantly influenced by the order in which the input is read, which can be a drawback in processing complex document layouts.

This realization steered us towards adopting an LMDX\cite{perot2023lmdx}-like method. Our choice of using an LMDX-like method was motivated by its generative capabilities, which we believe offer a more flexible and effective approach for the tasks at hand. We proceeded to fine-tune the Mistral-7B model \cite{jiang2023mistral} to produce text outputs that align closely with the ground truth data. This fine-tuning process was executed on a single A100 GPU and spanned over a 24-hour timeframe. The insights gleaned from this exercise are encapsulated in the accompanying table, providing a clear overview of our findings. Our initial baseline results for the key-value pair detection task are given in Table \ref{tab:baseline}

\begin{table}[ht]
\centering
\caption{Baseline Results for the key-value pair detection task}
\label{tab:baseline}
\begin{tabular}{lccccccccc}
\toprule
& \multicolumn{3}{c}{Text Only} & \multicolumn{3}{c}{Location Only} & \multicolumn{3}{c}{Text + Location (All)} \\
\cmidrule(r){2-4} \cmidrule(lr){5-7} \cmidrule(l){8-10}
& Precision & Recall & F1 & Precision & Recall & F1 & Precision & Recall & F1 \\
\midrule
Regular & 0.678 & 0.641 & 0.659 & 0.670 & 0.631 & 0.650 & 0.627 & 0.595 & 0.611 \\
Unkeyed & 0.584 & 0.620 & 0.601 & 0.635 & 0.672 & 0.653 & 0.568 & 0.601 & 0.584 \\
Unvalued & 0.617 & 0.586 & 0.601 & 0.634 & 0.604 & 0.618 & 0.603 & 0.573 & 0.588 \\
All & 0.645 & 0.640 & 0.643 & 0.665 & 0.657 & 0.661 & 0.615 & 0.608 & 0.612 \\
\bottomrule
\end{tabular}
\end{table}
\section{Conclusions}

In conclusion, the task of extracting information from business documents, especially without the crutch of predefined keys, presents a formidable challenge that spans across various domains. Our work sheds light on the critical gap within current datasets and benchmarks that are largely tailored for KIE with predetermined keys. By introducing \NAMEOFDATASET, not only do we provide a resource that caters to the nuanced demands of non-predetermined KVP extraction, we also set a new precedent for the depth of diversity and annotation detail required for meaningful progress in this field.

The significance of our contribution lies not just in the dataset itself but also in the potential it unlocks for future research and applications. By offering a platform that is both challenging and reflective of real-world complexities, \NAMEOFDATASET invites a broader exploration of methodologies and technologies in the realm of information extraction. This, in turn, could catalyze a wave of innovations that enhance the efficiency and accuracy of processing complex business documents.

As the community engages with \NAMEOFDATASET, we anticipate the emergence of novel approaches that not only excel in the context of our benchmark but also inspire the development of more adaptive, robust solutions for information extraction at large. Thus, our work not only addresses an immediate need within the field but also lays the groundwork for ongoing advancements that could redefine the boundaries of what's possible in information extraction from business documents.

\printbibliography

@inproceedings{ding2023form,
  title={Form-NLU: Dataset for the Form Natural Language Understanding},
  author={Ding, Yihao and Long, Siqu and Huang, Jiabin and Ren, Kaixuan and Luo, Xingxiang and Chung, Hyunsuk and Han, Soyeon Caren},
  booktitle={Proceedings of the 46th International ACM SIGIR Conference on Research and Development in Information Retrieval},
  pages={2807--2816},
  year={2023}
}

@inproceedings{yang2023modelingSIBR,
  title={Modeling Entities as Semantic Points for Visual Information Extraction in the Wild},
  author={Yang, Zhibo and Long, Rujiao and Wang, Pengfei and Song, Sibo and Zhong, Humen and Cheng, Wenqing and Bai, Xiang and Yao, Cong},
  booktitle={Proceedings of the IEEE/CVF Conference on Computer Vision and Pattern Recognition},
  pages={15358--15367},
  year={2023}
}

@inproceedings{jaume2019funsd,
  title={Funsd: A dataset for form understanding in noisy scanned documents},
  author={Jaume, Guillaume and Ekenel, Hazim Kemal and Thiran, Jean-Philippe},
  booktitle={2019 International Conference on Document Analysis and Recognition Workshops (ICDARW)},
  volume={2},
  pages={1--6},
  year={2019},
  organization={IEEE}
}

@inproceedings{huang2019icdar2019SORIE,
  title={Icdar2019 competition on scanned receipt ocr and information extraction},
  author={Huang, Zheng and Chen, Kai and He, Jianhua and Bai, Xiang and Karatzas, Dimosthenis and Lu, Shijian and Jawahar, CV},
  booktitle={2019 International Conference on Document Analysis and Recognition (ICDAR)},
  pages={1516--1520},
  year={2019},
  organization={IEEE}
}

@inproceedings{park2019cord,
  title={CORD: a consolidated receipt dataset for post-OCR parsing},
  author={Park, Seunghyun and Shin, Seung and Lee, Bado and Lee, Junyeop and Surh, Jaeheung and Seo, Minjoon and Lee, Hwalsuk},
  booktitle={Workshop on Document Intelligence at NeurIPS 2019},
  year={2019}
}

@inproceedings{stanislawek2021kleister,
  title={Kleister: key information extraction datasets involving long documents with complex layouts},
  author={Stanis{\l}awek, Tomasz and Grali{\'n}ski, Filip and Wr{\'o}blewska, Anna and Lipi{\'n}ski, Dawid and Kaliska, Agnieszka and Rosalska, Paulina and Topolski, Bartosz and Biecek, Przemys{\l}aw},
  booktitle={International Conference on Document Analysis and Recognition},
  pages={564--579},
  year={2021},
  organization={Springer}
}

@inproceedings{wang2023vrdu,
  title={Vrdu: A benchmark for visually-rich document understanding},
  author={Wang, Zilong and Zhou, Yichao and Wei, Wei and Lee, Chen-Yu and Tata, Sandeep},
  booktitle={Proceedings of the 29th ACM SIGKDD Conference on Knowledge Discovery and Data Mining},
  pages={5184--5193},
  year={2023}
}

@inproceedings{xu2022xfund,
  title={Xfund: A benchmark dataset for multilingual visually rich form understanding},
  author={Xu, Yiheng and Lv, Tengchao and Cui, Lei and Wang, Guoxin and Lu, Yijuan and Florencio, Dinei and Zhang, Cha and Wei, Furu},
  booktitle={Findings of the Association for Computational Linguistics: ACL 2022},
  pages={3214--3224},
  year={2022}
}

@inproceedings{huang2022layoutlmv3,
  title={Layoutlmv3: Pre-training for document ai with unified text and image masking},
  author={Huang, Yupan and Lv, Tengchao and Cui, Lei and Lu, Yutong and Wei, Furu},
  booktitle={Proceedings of the 30th ACM International Conference on Multimedia},
  pages={4083--4091},
  year={2022}
}

@article{lee2023formnetv2,
  title={FormNetV2: Multimodal Graph Contrastive Learning for Form Document Information Extraction},
  author={Lee, Chen-Yu and Li, Chun-Liang and Zhang, Hao and Dozat, Timothy and Perot, Vincent and Su, Guolong and Zhang, Xiang and Sohn, Kihyuk and Glushnev, Nikolai and Wang, Renshen and others},
  journal={arXiv preprint arXiv:2305.02549},
  year={2023}
}

@article{hong2020bros,
  title={Bros: A pre-trained language model for understanding texts in document},
  author={Hong, Teakgyu and Kim, DongHyun and Ji, Mingi and Hwang, Wonseok and Nam, Daehyun and Park, Sungrae},
  year={2020}
}

@inproceedings{mathur2023layerdoc,
  title={LayerDoc: Layer-wise Extraction of Spatial Hierarchical Structure in Visually-Rich Documents},
  author={Mathur, Puneet and Jain, Rajiv and Mehra, Ashutosh and Gu, Jiuxiang and Dernoncourt, Franck and Tran, Quan and Kaynig-Fittkau, Verena and Nenkova, Ani and Manocha, Dinesh and Morariu, Vlad I and others},
  booktitle={Proceedings of the IEEE/CVF Winter Conference on Applications of Computer Vision},
  pages={3610--3620},
  year={2023}
}

@article{wang2020docstruct,
  title={Docstruct: A multimodal method to extract hierarchy structure in document for general form understanding},
  author={Wang, Zilong and Zhan, Mingjie and Liu, Xuebo and Liang, Ding},
  journal={arXiv preprint arXiv:2010.11685},
  year={2020}
}

@article{hwang2020spatial,
  title={Spatial dependency parsing for semi-structured document information extraction},
  author={Hwang, Wonseok and Yim, Jinyeong and Park, Seunghyun and Yang, Sohee and Seo, Minjoon},
  journal={arXiv preprint arXiv:2005.00642},
  year={2020}
}

@article{beltagy2020longformer,
  title={Longformer: The long-document transformer},
  author={Beltagy, Iz and Peters, Matthew E and Cohan, Arman},
  journal={arXiv preprint arXiv:2004.05150},
  year={2020}
}

@article{liu2019roberta,
  title={Roberta: A robustly optimized bert pretraining approach},
  author={Liu, Yinhan and Ott, Myle and Goyal, Naman and Du, Jingfei and Joshi, Mandar and Chen, Danqi and Levy, Omer and Lewis, Mike and Zettlemoyer, Luke and Stoyanov, Veselin},
  journal={arXiv preprint arXiv:1907.11692},
  year={2019}
}

@inproceedings{wang2021towards,
  title={Towards robust visual information extraction in real world: New dataset and novel solution},
  author={Wang, Jiapeng and Liu, Chongyu and Jin, Lianwen and Tang, Guozhi and Zhang, Jiaxin and Zhang, Shuaitao and Wang, Qianying and Wu, Yaqiang and Cai, Mingxiang},
  booktitle={Proceedings of the AAAI Conference on Artificial Intelligence},
  volume={35},
  number={4},
  pages={2738--2745},
  year={2021}
}

@article{perot2023lmdx,
  title={LMDX: Language Model-based Document Information Extraction and Localization},
  author={Perot, Vincent and Kang, Kai and Luisier, Florian and Su, Guolong and Sun, Xiaoyu and Boppana, Ramya Sree and Wang, Zilong and Mu, Jiaqi and Zhang, Hao and Hua, Nan},
  journal={arXiv preprint arXiv:2309.10952},
  year={2023}
}

@inproceedings{mathew2021docvqa,
  title={Docvqa: A dataset for vqa on document images},
  author={Mathew, Minesh and Karatzas, Dimosthenis and Jawahar, CV},
  booktitle={Proceedings of the IEEE/CVF winter conference on applications of computer vision},
  pages={2200--2209},
  year={2021}
}

@article{naparstek2022businet,
  title={BusiNet--a Light and Fast Text Detection Network for Business Documents},
  author={Naparstek, Oshri and Azulai, Ophir and Rotman, Daniel and Burshtein, Yevgeny and Staar, Peter and Barzelay, Udi},
  journal={arXiv preprint arXiv:2207.01220},
  year={2022}
}

@inproceedings{smith2007overview,
  title={An overview of the Tesseract OCR engine},
  author={Smith, Ray},
  booktitle={Ninth international conference on document analysis and recognition (ICDAR 2007)},
  volume={2},
  pages={629--633},
  year={2007},
  organization={IEEE}
}

@article{jiang2023mistral,
  title={Mistral 7B},
  author={Jiang, Albert Q and Sablayrolles, Alexandre and Mensch, Arthur and Bamford, Chris and Chaplot, Devendra Singh and Casas, Diego de las and Bressand, Florian and Lengyel, Gianna and Lample, Guillaume and Saulnier, Lucile and others},
  journal={arXiv preprint arXiv:2310.06825},
  year={2023}
}
\appendix
\section{Annotation format}
\label{appendix_annotation}
An exaple for flat KVP annotation is given in listing \ref{lst:flat_kvp}.
An exaple for unkeyed value annotation is given in listing \ref{lst:unkeyed_value} and an example with a section is given in \ref{lst:kvp_section}.
Note that in the annotation the $x$ any $y$ coordinates are relative to the page size and take values between $0$ and $1$
\begin{lstlisting}[float,language=json,style=json-style, breaklines=true, frame=single, caption={Example of Annotation Format}, label=lst:flat_kvp]
{
"rectangles": [
{
 "_id": "3ba88dfc-aee5-433b-a68d-a2616033cbd3",
 "annotationQCEvaluations": null,
 "annotationQcEvals": null,
 "attributes": {},
 "color": "rgb(253, 255, 0)",
 "comments": [],
 "coordinates": [
   {
     "x": 0.657255,
     "y": 0.4894
   },
   {
     "x": 0.66902,
     "y": 0.4894
   },
   {
     "x": 0.66902,
     "y": 0.542701
   },
   {
     "x": 0.657255,
     "y": 0.542701
   }
 ],
 "evaluations": null,
 "label": "Text",
 "label_id": null,
 "origin": "manual",
 "state": "editable",
 "type": ""
},

}
\end{lstlisting}

\begin{lstlisting}[float,language=json,style=json-style, breaklines=true, frame=single, caption={Example of Annotation Format}, label=lst:unkeyed_value]
{
"rectangles": [
{
 "_id": "deaed54b-ef4c-4152-a916-94715312ad11",
 "annotationQCEvaluations": null,
 "annotationQcEvals": null,
 "attributes": {},
 "color": "rgb(142, 165, 183)",
 "comments": [],
 "coordinates": [
   {
     "x": 0.171097,
     "y": 0.393928
   },
   {
     "x": 0.312631,
     "y": 0.393928
   },
   {
     "x": 0.312631,
     "y": 0.408258
   },
   {
     "x": 0.171097,
     "y": 0.408258
   }
 ],
 "evaluations": null,
 "label": "Floating Address",
 "label_id": null,
 "origin": "manual",
 "state": "editable",
 "type": ""
},
]
}
\end{lstlisting}
\begin{lstlisting}[float,language=json,style=json-style, breaklines=true, frame=single, caption={Example of Annotation Format}, label=lst:kvp_section]
{
"rectangles": [
 {
   "_id": "5a331350-94d0-4dad-b5db-06e992d0a902",
   "annotationQCEvaluations": null,
   "annotationQcEvals": null,
   "attributes": {},
   "color": "rgb(216, 82, 82)",
   "comments": [],
   "coordinates": [
     {
       "x": 0.084706,
       "y": 0.66808
     },
     {
       "x": 0.213333,
       "y": 0.66808
     },
     {
       "x": 0.213333,
       "y": 0.685645
     },
     {
       "x": 0.084706,
       "y": 0.685645
     }
   ],
   "evaluations": null,
   "label": "Section",
   "label_id": null,
   "origin": "manual",
   "state": "editable",
   "type": ""
 },

}
\end{lstlisting}
\section{Additional Statistics}
\begin{figure}[htbp]
\centering
\includegraphics[width=1\columnwidth]{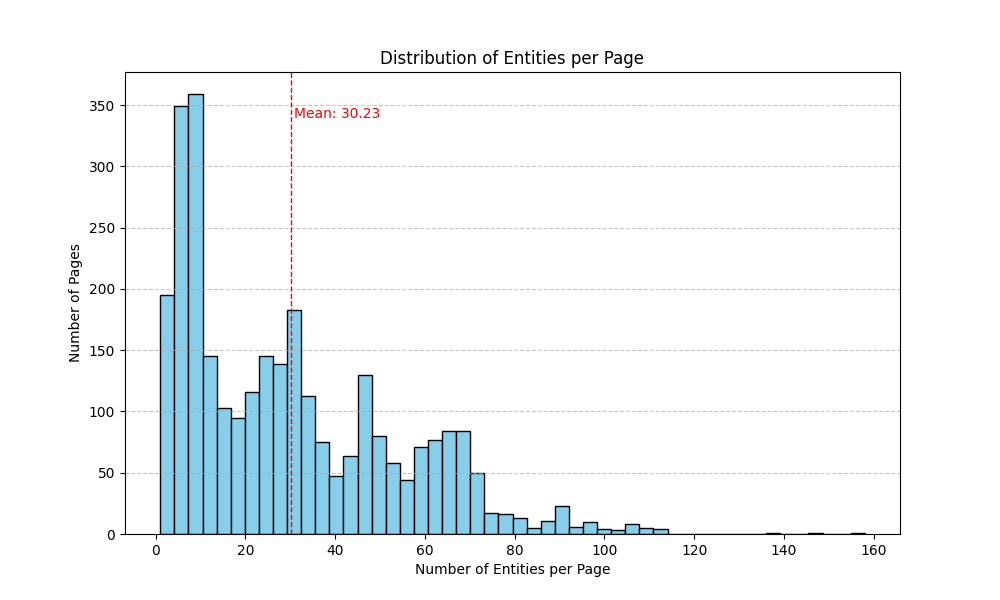}
\caption{Distribution of entities per page in \NAMEOFDATASET sourced from Public files.}
\label{fig:publicfiles_entities_per_page_histogram}
\end{figure}

\begin{figure}[htbp]
\centering
\includegraphics[width=1\columnwidth]{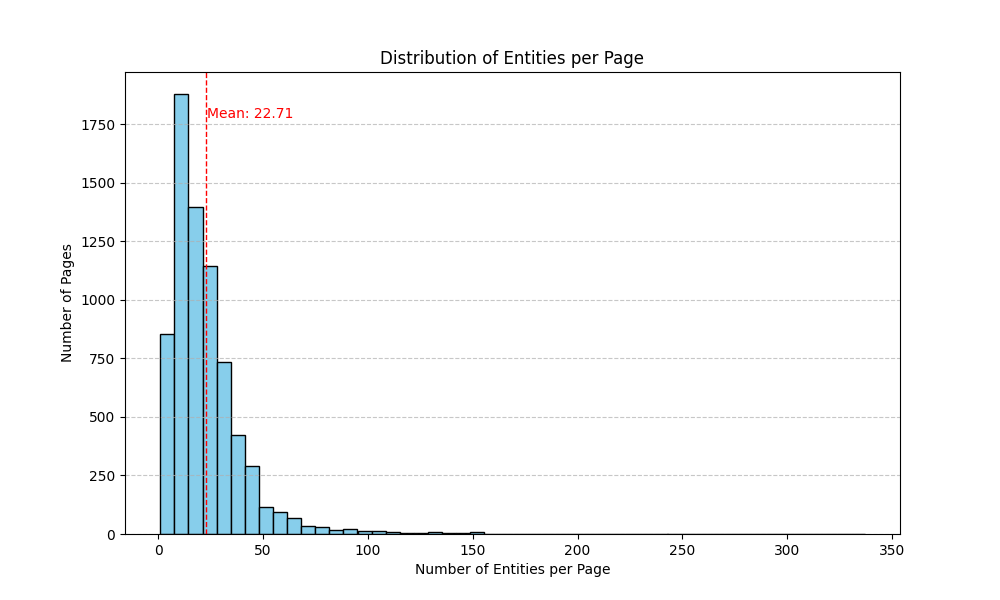}
\caption{Distribution of entities per page in \NAMEOFDATASET sourced from web crawl.}
\label{fig:zrl_entities_per_page_histogram}
\end{figure}

Here we provide mode detailed statistics of the data.
Figure \ref{fig:publicfiles_entities_per_page_histogram} is a histogram that specifically focuses on documents sourced from public files. 
Figure \ref{fig:zrl_entities_per_page_histogram}, on the other hand, is a histogram that specifically focuses on documents sourced from web crawl. 

These plots provides an overview of the variability in the number of entities in total, including key-value pairs and other types of entities, across different documents and within specific sources. The aim is to offer a comprehensive view of the dataset's characteristics, highlighting variations in document complexity across different sources.

\begin{figure}[htbp]
\centering
\includegraphics[width=1\columnwidth]{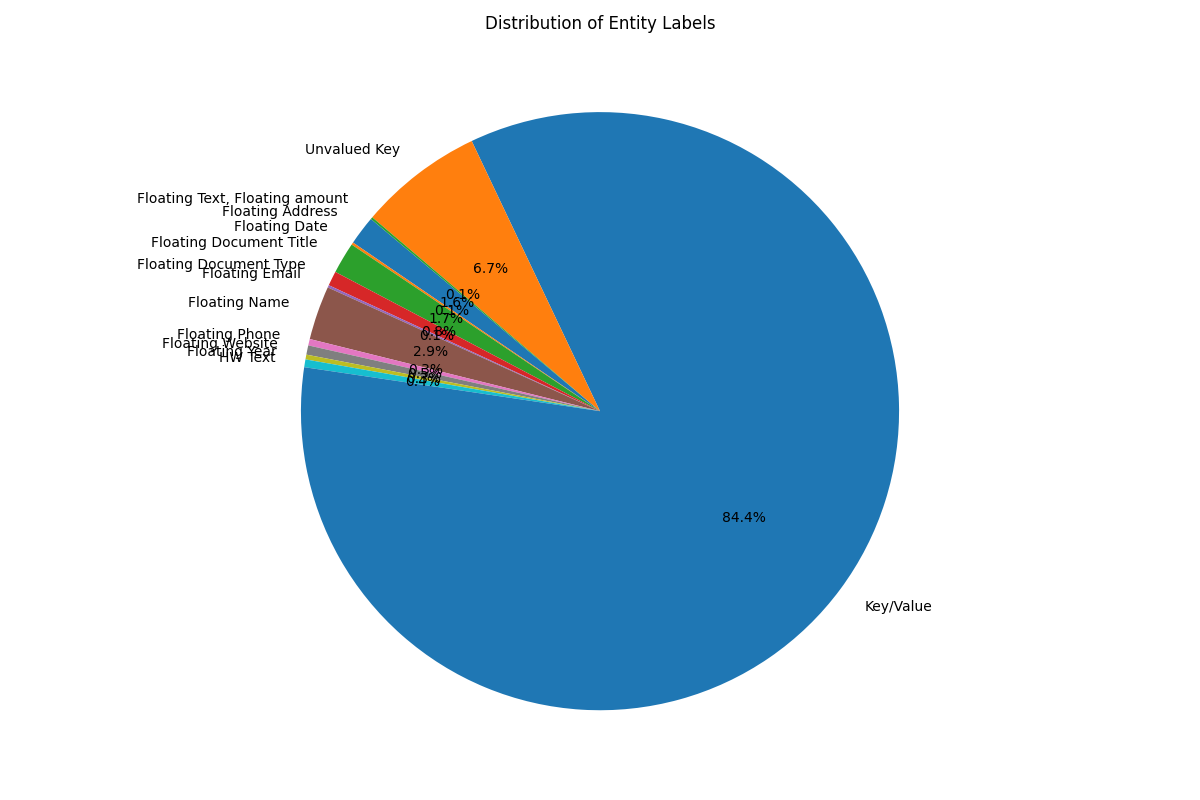}
\caption{Distribution of entity labels in \NAMEOFDATASET sourced from Public files.}
\label{fig:publicfiles_entity_labels_pie_chart}
\end{figure}

In addition to the overall distribution, we have generated two more pie charts to examine the distribution of entity labels within specific subsets of the dataset:

- Figure \ref{fig:publicfiles_entity_labels_pie_chart} focuses on documents sourced from public files. 

\begin{figure}[htbp]
\centering
\includegraphics[width=1\columnwidth]{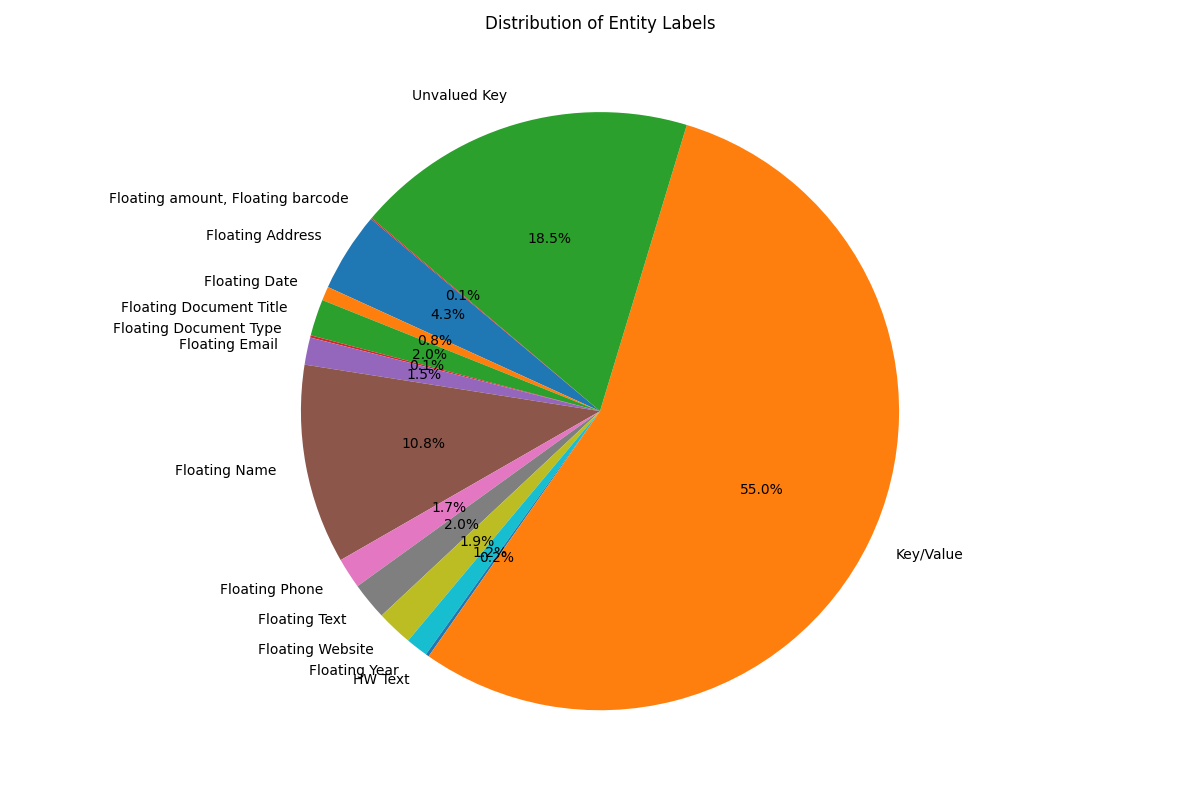}
\caption{Distribution of entity labels in \NAMEOFDATASET sourced from web crawl.}
\label{fig:zrl_entity_labels_pie_chart}
\end{figure}

- Figure \ref{fig:zrl_entity_labels_pie_chart} specifically examines documents sourced from web crawl. 

The following pie charts offer a more intuitive representation of the distribution of entity labels in total across different documents and within specific sources, facilitating a better understanding of the characteristics of the dataset.

\end{document}